\documentclass[reprint,
superscriptaddress,
 amsmath,amssymb,
 aps,pra
]{revtex4-2}
\usepackage{graphicx}
\usepackage{dcolumn}
\usepackage{bm}
\usepackage{xcolor}
\usepackage{ulem}

\DeclareMathOperator{\Real}{Re}
\begin{document}

\preprint{APS/123-QED}

\title{Optimal displacement detection of arbitrarily-shaped levitated dielectric objects using optical radiation}%



\author{Shaun Laing}
\affiliation{Physics Department, Faculty of Science and Engineering, Swansea University, Swansea SA2 8PP, UK}
\author{Shelby Klomp}
\author{George Winstone}
\author{Alexey Grinin}
\author{Andrew Dana}
\author{Zhiyuan Wang}
\author{Kevin Seca Widyatmodjo}
\affiliation{Center for Fundamental Physics, Department of Physics and Astronomy,Northwestern University, Evanston, Illinois 60208, USA}
\author{James Bateman}
\affiliation{Physics Department, Faculty of Science and Engineering, Swansea University, Swansea SA2 8PP, UK}
\author{Andrew A. Geraci}
\affiliation{Center for Fundamental Physics and Center for Interdisciplinary Exploration and Research in Astrophysics, Department of Physics and Astronomy, Northwestern University, Evanston, Illinois 60208, USA}

\date{\today}

\begin{abstract}

Optically-levitated dielectric objects are promising for precision force, acceleration, torque, and rotation 
sensing due to their extreme environmental decoupling. While many levitated optomechanics experiments employ spherical objects, for some applications non-spherical geometries offer advantages. For example, rod-shaped or dumbbell shaped particles have been demonstrated for torque and rotation sensing and high aspect ratio plate-like particles can exhibit reduced photon recoil heating and may be useful for high-frequency gravitational wave detection 
or as 
high bandwidth accelerometers. To achieve optimal sensitivity, cooling, and quantum control in these systems, it is beneficial to achieve  optimal displacement detection using scattered light. We describe and numerically implement a method based on Fisher information that is applicable to suspended particles of arbitrary geometry. We demonstrate the agreement between our method and prior methods employed for spherical particles, both in the Rayleigh and Lorentz-Mie regimes. 
As practical examples, we analyze the optical detection limits of an optically-levitated high-aspect-ratio disc-like dielectric object and a rod-shaped object for configurations recently realized in experimental work.
\end{abstract}

%



\maketitle

\section{Introduction}
Optimal position and momentum measurements are important for many applications of optical tweezers. In life- and material sciences, efficient detection allows for lower laser powers and thus less perturbation of the object under study \cite{Bustamante2021,NEUMAN19992856}. 
For precision measurements in quantum-optomechanical systems, detection efficiency also impacts displacement sensitivity due to shot noise and radiation pressure back-action \cite{optomechreview} and determines the lowest temperature achievable for the center-of-mass motion of the suspended particle when applying feedback cooling. 
Levitated optomechanics, where an object can be suspended with radiation pressure, diamagnetic levitation, or radio-frequency electromagnetic fields, is a platform that features excellent decoupling from the environment, leading to a number of applications in precision sensing, quantum science, materials science, and non-equilibrium physics. For recent reviews see Refs. \cite{Millen_2020,reviewmarkus,moore2021searching}.
The information that can be attained regarding the displacement of a levitated particle via scattering of optical radiation has been quantified in recent literature using the information radiation pattern (IRP) \cite{tebbenjohanns2019optimal,maurer2022quantum,Maurer2024}. The information radiation pattern will generally depend on the observation direction, the shape of the particle, and the detection method. 

The theory of optimal position detection for the Rayleigh regime has been studied in \cite{tebbenjohanns2019optimal} and for Mie-spheres in \cite{maurer2022quantum,Maurer2024}. While dipolar scatterers and Mie-spheres are frequently encountered cases, particles of other shapes, particularly high-aspect-ratio disks and rods, dumbbells, and hollow spheres, are optimal for many applications.
For instance, rod-shaped or dumbbell shaped particles are well-suited for sensitive torque and rotation sensing \cite{Li2018,Novotny2018,Moore2018,Xu2017Detecting} and high aspect ratio plate-like particles can exhibit reduced photon recoil heating for high bandwidth accelerometers or the detection of high-frequency gravitational waves with frequencies exceeding a few kHz \cite{aggarwal2022searching, winstone2022optical}. It is therefore of interest to develop a method to provide a practical means to calculate IRPs for experimental configurations which include scatterers of arbitrary shape. Furthermore, it is of practical interest to consider how realistic detection schemes can approach the ideal theoretical limits for \textcolor{black}{detection efficiency} for useful levitated particle geometries.

\textcolor{black}{Although other approximate analytic solutions exist for scattering from dielectric particles with certain properties and geometries, (e.g., the Rayleigh--Gans approximation for optically soft particles, or e.g. as described in Ref. \cite{lightscattering1979} for dielectric bodies which are small compared to the wavelength in at least one direction), the range of validity of such expressions does not adequately span the geometries with the aforementioned applications of interest. Additionally, while recent progress has been made on the challenging problem of  establishing approximate analytic expressions for scattering by non-spherical particles in the Mie regime e.g. for prolate or oblate spheroids in certain limiting cases \cite{Ding:23}, the need remains to understand the scattering and information radiation patterns of objects of a wider variety of shapes and sizes.} 
\textcolor{black}{In Ref. \cite{Rotter2024}, a method for in principle determining the information radiation pattern of dielectric objects of arbitrary geometry was described based on the flow of Fisher information in wave scattering.}
In this paper we present \textcolor{black}{and benchmark} a numerical method for determining the optimal interferometric position measurement for arbitrarily-shaped dielectric objects and derive equations for their IRP, based on the Fisher information.  We demonstrate reasonable agreement with the methods shown in Refs. \cite{tebbenjohanns2019optimal} and \cite{maurer2022quantum,Maurer2024} for spherical particles, while generalizing it to arbitrary shapes. 
We demonstrate several numerical implementations using the SCUFF-EM package \cite{Reid2012,Reid2018}, the pyGDM package \cite{pyGDM2,wiecha2022pygdm}, and the COMSOL Multiphysics \cite{comsol} package to determine the IRPs for disks, rods, and spheres.  
In addition to the center of mass degrees of freedom corresponding to translational motion, we analyze and discuss \textcolor{black}{librational} degrees of freedom of non-spherical objects. 

This paper is structured as follows. In Sec. \ref{sec:theory} we describe the theoretical approach being used for determining IRP based on Fisher information. In Sec. \ref{sec:methods} we describe its numerical implementation in several computational packages and conduct a comparison of the results. In Sec. \ref{sec:results} we present and discuss numerical IRPs in a focused Gaussian beam trap and counter propagating beam trap for an ideal reference field for spheres, plates, and rods. For the case of Mie scattering by spheres, we show the equivalence between our results and those in Ref. \cite{maurer2022quantum}.  In section \ref{sec:reference}, we present an example of numerical results for lab-relevant Gaussian reference fields and 
consider detector geometries for practical detection 
configurations. 
Finally we conclude with a discussion of future applications of relevance to the optomechanics community, including the detection of high frequency gravitational waves.



\section{Theoretical approach}
\label{sec:theory}





In Ref. \cite{maurer2022quantum} the information radiation pattern for light scattered by Mie spheres was calculated by considering the quantum theory of excitation of exact eigenmodes of the electromagnetic field in the presence of the dielectric sphere at equilibrium.  To compute the IRP for a scatterer of generalized geometry, we adopt the method of computing the Fisher information in the scattered radiation field \cite{oriolprivate,Rotter2024}. A similar approach was recently used to determine the information radiation pattern for reflective geometries \cite{Bateman2024}.  

\textcolor{black}{The Fisher information flux carried by an electromagnetic wave scattered in the far field ($\theta,\phi$) can be expressed as}
\begin{equation}
    \vec{S}^{\text{FI}}(\theta,\phi)_\mu = 
    \frac{2}{\hbar\omega}
    \Real\left(\frac{\partial }{\partial \mu}\vec{E}^*(\theta,\phi)
    \times
    \frac{\partial}{\partial \mu}\vec{H}(\theta,\phi)\right),
    \label{EQ:Fisher Flux}
\end{equation}
where $\vec{E}$ and $\vec{H}$ are the electric and magnetic fields of the scattered radiation, respectively, \textcolor{black}{$\omega$ is the angular frequency of the incoming field}, and $\mu \in (x,y,z,\theta_x,\theta_y,\theta_z$) is the coordinate along which the motion of the trapped particle is being described.

\section{Numerical methods 
\label{sec:methods}}

In this section we describe the numerical implementation of generating scattered electric fields using three different computational back-ends and how these fields are used to calculate information radiation patterns with the Fisher Information method. Numerical methods are required in order to allow for calculation of IRPs for any arbitrary particle shape or incident field that may be difficult to compute analytically. We also include a detailed comparison of the relative advantages and limitations of each computational package to allow people who wish to apply these methods for their own use to choose an appropriate back-end for their own purposes.


SCUFF-EM (\textbf{S}urface \textbf{CU}rrent/\textbf{F}ield \textbf{F}ormulation of \textbf{E}lectro\textbf{M}agnetism) is an open-source implementation of the surface integral equation/boundary-element method (SIE/BEM) to solve electromagnetic scattering problems~\cite{Reid2015,Reid2018}. Starting with a known geometry in some medium (this could be a vacuum, a fluid, or a gas) and illuminating the geometry with some known incident EM field, we can use these techniques to find the unknown scattered field. As a numerical solver, the scatterer must be discretized. We mesh the object we are interested in using a set of roughly equal triangles. Each interior edge of the mesh is given a basis function $\vec{f}_n(\vec{r})$ which represents the surface currents~\cite{Wilton1980}. These basis functions are used to solve the following linear system,
\begin{equation}
    \langle \vec{f}_m|\mathbf{\Gamma}^{EE}|\vec{f}_n \rangle \vec{k}
    = - \langle\vec{f}_m|\vec{E}^\text{inc}\rangle
    \label{EQ:SCUFF-EM equation}
\end{equation}
where $\mathbf{\Gamma}^{AB}$ is the $3\times 3$ Dyadic Green's function giving the field of type $A$ from the surface current of type $B$, $\vec{E}^\text{inc}$ is the incident electric field. \textcolor{black}{The elements of the vector $\vec{k}$ are the coefficients $k_n$ of the surface-current basis functions $\vec{f}_n$.}
The scattered electric and magnetic fields at a point $\vec{r}$ are given by,
\begin{align}
    \vec{E}_s(\vec{r}) &= 
    \sum_n k_n \int \mathbf{\Gamma}^{EE}(\vec{r},\vec{r}')\vec{f}_n(\vec{r'})d\vec{r}'
    \label{EQ:SCUFF-EM electric field}\\
    \vec{H}_s(\vec{r}) &= 
    \sum_n k_n \int \mathbf{\Gamma}^{ME}(\vec{r},\vec{r}')\vec{f}_n(\vec{r'})d\vec{r}'
    \label{EQ:SCUFF-EM magnetic field}
\end{align}
respectively~\cite{Reid2012}. 

Another numerical method we use is pyGDM2\cite{ wiecha2022pygdm, Girard_2005}, which is the Python toolkit for full-field electro-dynamical simulations of nano-structures based on the Green Dyadic Method (GDM). Similar to the coupled dipole approximation, pyGDM calculates the optical response of an arbitrary shaped object by discretizing the volume of the object into a sum over finite size volume elements the in Lippmann-Schwinger equation 

\begin{align}
    \vec{E}(\vec{r_i}) &= \vec{E}^\text{inc}(\vec{r_i})+
    \sum_{j=1}^{N} \mathbf{S}_{0}^{EE}(\vec{r_i},\vec{r_j})\cdot \mathbf{\chi}_e(\vec{r_j})\cdot \vec{E}(\vec{r_j})V_{cell}
\end{align}
where $\vec{E}^\text{inc}$ is the incident electric field, $\mathbf{S}_{0}^{EE}$ is the field-susceptibility tensor, which is closely related to the Dyadic Green's function $\mathbf{\Gamma}^{EE}$ ~\cite{PhysRevA.11.230}; $\mathbf{\chi}_e$ is the susceptibility of the structure, and $\vec{r_i}$ is the position of each volume element $V_{cell}$. Then one can construct a 3N-dimensional vector to represent the ensemble of all the electric field of the finite volume element for the incident field or the scattered field $\vec{E}^\text{inc}_{obj}$ or $\vec{E}_{obj}$ and use a generalized field propagator $\mathbf{K}$, which is a $3N\times3N$
matrix describing the response of the entire structure. Once the generalized propagator $\mathbf{K}$ is known, we can calculate the response of the system to arbitrary monochromatic incident fields by a matrix-vector,
\begin{align}
\vec{E}_{obj} = \mathbf{K}\cdot\vec{E}^\text{inc}_{obj}.
\end{align}

In addition to SCUFF-EM and pyGDM, we also use COMSOL Multiphysics which is a simulation software that uses the finite element method to solve partial differential equations relevant to many physics problems. In this case, the \textit{Wave Optics Module} is used to simulate scattered electric fields within the \textit{Electromagnetic Waves, Frequency Domain} Interface. Similarly to our other methods, the known geometry and incident EM fields must be defined, but in COMSOL a perfectly matched layer is implemented to create the correct boundary conditions at the edges of the simulation, and the volume of the scatterer and the surrounding volume is discretized. The mesh is determined by the physics of the problem within COMSOL to be primarily tetrahedral with a minimum of 5 mesh elements per wavelength. The solver computes the near field scattered fields in the entire geometry by iteratively computing Maxwell's equations for all mesh points. The calculation relies on a biconjugate gradient stabilized iterative method \cite{comsol}.  Then the far field electric field is calculated from the near field on a particular surface $S$ enclosing the particle, by using the Stratton-Chu formula,
\begin{equation}
    \vec{E}_{\mathrm{far}}(\theta,\phi) = 
    \frac{i k}{4 \pi}\vec{r}_0\times
    \int [\vec{n}\times\vec{E}-\eta\vec{r}_0\times(\vec{n}\times\vec{H})]
    e^{i k \vec{r}\cdot\vec{r}_0}dS
    \label{EQ:Stratton-Chu equation}
\end{equation}
where $\vec{r}_0$ is the unit vector in the direction of radius vector $\vec{r}$ which points from the origin to the surface $S$, $\vec{n}$ is the unit normal to the surface $S$, and $\eta$ is the impedance $\eta=\sqrt{\mu/\epsilon}$.


\begin{table}
    \centering
    \begin{tabular}{cccc}
        \hline \hline
        Package & Mesh Topology & Scaling & Method\\
        \hline
        SCUFF-EM & Area & $A^4$ & Boundary element \\
        PyGDM2 & Volume & $V^3$ & Green's Dyadic\\
        COMSOL & Volume & Adaptive & Variable\\
        \hline \hline
    \end{tabular}
    \caption{Comparison of numerical methods used to generate scattered electric fields.}
    \label{Table:Computational Methods}
\end{table}



The three methods that we used for the numerical calculation of the scattered fields are listed in Table \ref{Table:Computational Methods}, where we compare the factors that contribute to the efficiency of the method. The meshing in PyGDM and COMSOL is done for the volume of the relevant geometry, which can be computationally taxing for large geometries. In PyGDM, the volume of only the scatterer itself must be dicretized, and the mesh elements are equally sized throughout the geometry, so the scaling of computation time goes as the cube of the volume (i.e. the cube of the number of elements). COMSOL requires that the volume of the scatterer and an additional surrounding volume be discretized, but it has the advantage of adaptive meshing, where the physics of the problem is taken into consideration in the generation of the mesh. For example, the mesh is often finest on the particle surface, but has the minimum spacing of 5 mesh elements per wavelength farther away. Additionally, in the calculation of IRPs, it is important that the surface that is used to generate the far field in Eq. \ref{EQ:Stratton-Chu equation} has a sufficiently fine mesh. SCUFF-EM, on the other hand, uses methods where the calculation is done only on the surface of the particle, so the mesh is an evenly spaced triangular mesh, and the computation time scales with the surface area of the scatterer. For this reason, SCUFF-EM often has much shorter computation times for large particle geometries. 

Another important factor in the comparison of these methods is the solver type. Unlike PyGDM and SCUFF-EM which use fixed methods and solvers with different memory and CPU overheads\cite{pyGDM2, Reid2012}, COMSOL uses a selection of direct and iterative solvers that it will choose from based on the physics of the problem being solved. For the scattering problems discussed here, COMSOL primarily employed the biconjugate gradient stabilized (BiCGStab) iterative method \cite{comsol}.




In order to convert the numerically calculated scattered fields into IRPs we must measure how the field changes on some measurement sphere in the far field regime as the particle moves in one axis. We achieve this by measuring the scattered field some distance $r$ from the origin with the particle located at $-l$ and $+l$ where the distance $2l \ll \lambda$ such that the change in the field as we move from $-l$ to $+l$ is linear. We find the complex scattered electric and magnetic fields $\vec{E}^{\pm l}(\theta,\phi)$ and $\vec{H}^{\pm l}(\theta,\phi)$ at each point $(\theta,\phi)$ on the measurement sphere when the particle is located at either $-l$ or $+l$. This measurement sphere is made up of hypothetical detectors each covering a solid angle $d\Omega$. To find the Fisher information flux Eq.~\eqref{EQ:Fisher Flux}, we find each component of the change of each field as the first-order central-difference approximation to the derivative,
\begin{align}
    \frac{\partial E^*_j}{\partial \mu} &= 
    \frac{E^{*+}_j - E^{*-}_j}{l-(-l)}
    \label{EQ:electric field differental}\\
    \frac{\partial H_j}{\partial \mu} &= 
    \frac{H^{+}_j - H^{-}_j}{l-(-l)}
    \label{EQ:magnetic field differental}
\end{align}
%
%
%

where $j\in (x,y,z)$ is the component of the field in cartesian coordinates, and $\mu$ is the axis the object is moved through. By inserting Eqs.~\eqref{EQ:electric field differental} and~\eqref{EQ:magnetic field differental} into Eq.~\eqref{EQ:Fisher Flux}, we find the Fisher information flux, and thus we can write the information flowing through surface $S$ as\textcolor{black}{
\begin{equation}
    \mathcal{I}^\text{Fisher}_{\mu} = \int_S \vec{S}^\text{FI}(\theta,\phi) \cdot \hat{n}(\theta,\phi)  d\Omega,
    \label{EQ:IRP full surface}
\end{equation}
where $\hat{n}$ is the outward facing unit vector located at $(\theta,\phi)$. By discretizing our surface into a set of area elements $\{A_i\}$ that are each sufficiently small such that $\vec{S}^\text{FI}$ does not vary across the area element, we can find the information through the $i$th element as
\begin{equation}
    \mathcal{I}^\text{Fisher}_{\mu,i} = \vec{S}^\text{FI}(\theta,\phi) \cdot 
    \vec{A}_i,
    \label{EQ:IRP numerical}
\end{equation}
where $\vec{A}_i$ is the area vector associated with the area element $A_i$.}
\textcolor{black}{The units of $\mathcal{I}_{\mu}$ in Eq. \eqref{EQ:IRP numerical} are found to be Hz m$^{-2}$ when $\mu$ is a displacement, or Hz rad$^{-2}$ when $\mu$ is an angular degree of freedom. 
That is, $\mathcal{I}^\text{Fisher}_{\mu,i}$ is the amount of information that flows through the area $A_i$ per second for the change in the parameter $\mu$. For the results presented here, we normalize $\mathcal{I}^\text{Fisher}_{\mu,i}$ by dividing it by the total scattered information, $\sum_i\mathcal{I}^\text{Fisher}_{\mu,i}$. We denote this normalized form as $\bar{\mathcal{I}}^\text{Fisher}_{\mu,i}$.}

A direct comparison of IRPs calculated using the three back-ends described here are shown in the Appendix \ref{AppA} where we show that the radiation patterns generated with each computational method are in reasonable agreement.

\section{Results for spheres, plates, rods}
\label{sec:results}




Using the Fisher information approach with numerical solvers, we can calculate IRPs for novel experimentally relevant geometries. In this section, we show IRPs for hexagonal plates and cylindrical rods which are currently outside of the results possible with analytics. 
\begin{figure}
    \centering
    \includegraphics[width=0.44\textwidth]{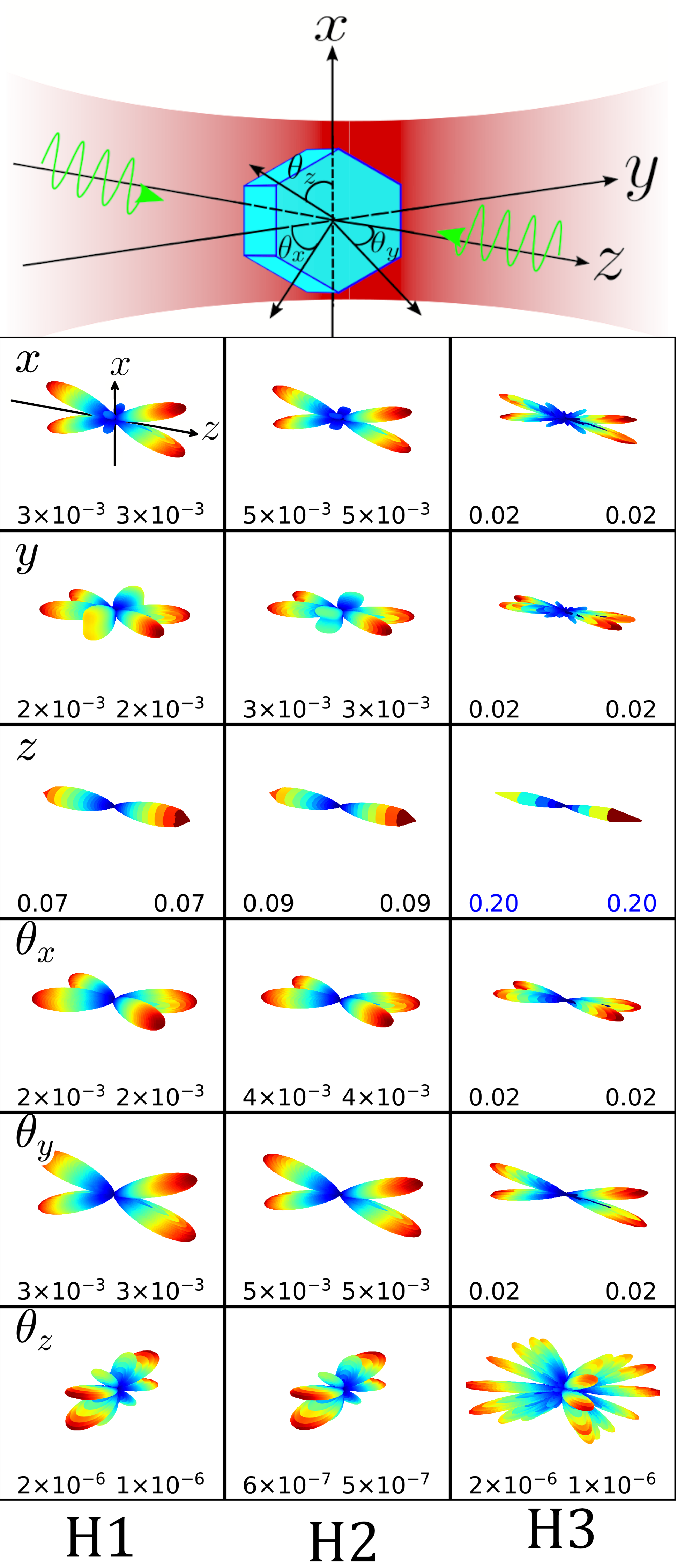}
    \caption{(upper) A hexagonal plate levitated in a standing wave optical trap and labeled coordinates corresponding to the displayed information radiation patterns. Laser fields are linearly polarized along the $x-$ direction and propagating along the $z-$ direction. Angles $\theta_x,\theta_y,\theta_z$ represent \textcolor{black}{librations} about the $x-,y-$ and $z-$ axes, respectively. (lower) IRPs from the SCUFF-EM implementation for each degree of freedom $\mu$ of a hexagonal disk trapped in a standing wave trap made up of two counter-propagating beams with wavelength $\lambda = 1550 ~\text{nm}$ and a beam waist of $w_0 = 12~\mu\text{m}$. Each plate has dimensions given in Table~\ref{Table:Hex Sizes}. We also show the detection efficiencies for the left and right focusing lenses each with a numerical aperture of $\text{NA} = 0.082$.}
    \label{Fig:Hex diagram}
\end{figure}
\begin{table}
    \centering
    \begin{tabular}{c c c}
        \hline \hline
        Batch Name & Thickness & Diameter\\
        \hline
        H1 & $300\text{ nm}$ & $2.5$ $\mu\text{m}$ \\
        H2 & $200\text{ nm}$ & $3$ $\mu\text{m}$\\
        H3 & $200\text{ nm}$ & $5$ $\mu\text{m}$\\
        \hline \hline
    \end{tabular}
    \caption{Dimensions of the hexagonal plates considered in the text. These dimensions roughly correspond to those reported in recent experimental work in Ref. \cite{winstone2022optical}}.
    \label{Table:Hex Sizes}
\end{table}
Results for a hexagonal plate trapped in a standing wave similar to the experimental configuration described in Ref. ~\cite{winstone2022optical} are shown in Fig.~\ref{Fig:Hex diagram}. We consider three sizes of plates labelled H1, H2, and H3 with dimensions given in Table~\ref{Table:Hex Sizes}. The IRPs generated by these plates are given in Fig.~\ref{Fig:Hex diagram} for a laser wavelength of $\lambda = 1550 ~\text{nm}$ and a beam waist of $w_0 = 12~\mu\text{m}$. \textcolor{black}{We note that the orientation of the hexagonal plates in the laser beam in general influences the shape and structure of the IRPs, and in this paper we consider only detection of small displacements and librations around the trapping equilibrium position in the standing wave trap, in particular with the hexagonal face of the prism oriented along the axis of the laser standing wave.} We also show the detection efficiencies for a lens to the left and a lens to the right. These are given in the lower corners of each panel. 
%
%

The detection efficiencies are found by summing the information through the area elements that make up the total area of the collection lens as \textcolor{black}{
\begin{equation}
    \eta_{\mu} = \sum_{A_i \in S_d} \bar{\mathcal{I}}^\text{Fisher}_{\mu,i}
    \label{EQ:Detection efficencies}
\end{equation}
where $S_d$ is the solid angle covered by the collection lens.} Blue detection efficiencies indicate a value greater than $1/9$, which is necessary for ground state cooling via active feedback cooling, as discussed in Ref. \cite{tebbenjohanns2019optimal}.


\begin{figure}
    \centering
    \includegraphics[width=0.30\textwidth]{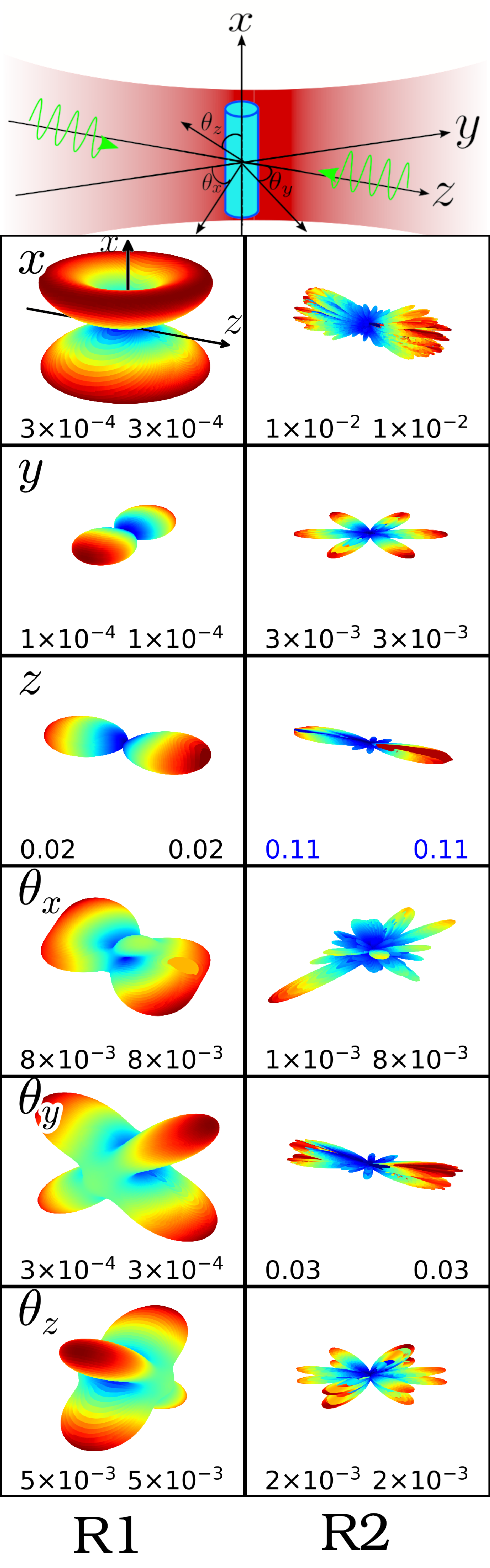}
    \caption{(upper) A rod levitated in a standing wave optical trap. Angles $\theta_x,\theta_y,\theta_z$ represent \textcolor{black}{librations} about the $x-,y-$ and $z-$ axes, respectively. (lower) IRPs from the SCUFF-EM implementation for each degree of freedom $\mu$ of a rod trapped in a standing wave trap made up of two counter-propagating beams with wavelength $\lambda = 1550 ~\text{nm}$ and a beam waist of $w_0 = 12~\mu\text{m}$. The rods have the dimensions given in Table~\ref{Table:Rod Sizes}. We also show the detection efficiencies for the left and right focusing lenses each with a numerical aperture of $\text{NA} = 0.082$.
    \label{Fig:Rod diagram}}
\end{figure}

We generally see that the information radiation pattern has a structure containing lobes. Assuming the laser propagates along $\hat{z}$, for the $x$ and $y$ motions, the main lobes are oriented in the $x-z$ and $y-z$ planes, respectively.  For motion in the $z$ axis the information is almost entirely concentrated in the $\pm z$ directions. This is unsurprising as the scattered radiation from lasers incident on the face of the flat plate tends to be emitted primarily in the forwards and backwards directions. \textcolor{black}{
The symmetry along the $\pm z$ axis results from the counter-propagating beams that form the trap. Symmetries in the other directions occur from the symmetry of the particle and its motion about the trapping minimum.  The structure of the lobes in general depends on the symmetry of the particle, the polarization of the detection laser, as well as the  dimensions of the particle in comparison to the detection laser wavelength.}
\textcolor{black}{For the hexagonal prisms the IRPs are generally mirror-symmetric about the x-z plane and the y-z plane. 
For translational motion, lobes generally appear in the direction of the particle's motion, although we note that the $x$ motion does not produce $x$ pointing lobes as no light is scattered in the direction of the incident beam polarization. 
We see no information along the $z$ axis for the $x$ and $y$ motion, as the phase and amplitude of the light scattered along $z$ has no first order dependence on the x- and y-translation when the particle is positioned near the z-axis. } 


\textcolor{black}{For a single beam, as shown also in the Appendix, the backwards scattered photons contain more information about the displacement along the laser beam axis. This agrees with the result for nanospheres in a single beam optical trap reported in Refs. \cite{tebbenjohanns2019optimal,maurer2022quantum}. Intuitively, this can be understood as the phase of forward transmitted light through the object does not vary with the position of the plate, whereas for backward scattered light, the phase seen on a detector depends on the location of the surface plate at the moment when reflection occurs. This is also consistent with the simple one-dimensional case of radiation impinging on a dielectric plate considered in Ref. \cite{Rotter2024} where the energy is mostly contained in the transmitted field while the position information is contained in the reflected field.}

The IRPs obtained for the \textcolor{black}{librational} degrees of freedom \textcolor{black}{can also be obtained in a straightforward manner with our method}. The $\theta_x$ and $\theta_y$ motion both produce lobes in the directions consistent with the direction light will primarily scatter into under \textcolor{black}{librations} about those axes. Lobes corresponding to $\theta_x$ and $\theta_y$ \textcolor{black}{librations} are visible in $y-z$ and $x-z$ planes, respectively. The $\theta_z$ motion primarily produces lobes radial to the laser propagation direction.  These result from the change in the angle of the radial faces on the hexagon and the detection efficiency for this motion is rather poor compared with \textcolor{black}{librations} about the other two axes, as expected. We also note that, in general, the detection efficiencies increase with the hexagon diameter. This corresponds with more of the IRP being concentrated forwards and backward as the diameter increases. 
However, in the $\theta_z$ \textcolor{black}{libration}, this is not the case. \textcolor{black}{We also note the detection efficiency for this motion is generally quite small, being primarily associated with the hexagonal edge corners.}
\textcolor{black}{
For the $\theta_z$ motion, we observe that the number of lobes in the IRP depends on hexagonal shape of the plate, as discussed further in Appendix \ref{AppB}. We find that for plate-like particles where the face is an even-sided polygon, the number of lobes appearing in $\theta_z$ scales with the number of sides (see Appendix).} 


Nanorods are another novel geometry that has gained much interest recently~\cite{Kuhn:17}. Trapping a rod in a standing wave will generate IRPs as shown in Fig.~\ref{Fig:Rod diagram}. \textcolor{black}{We consider the IRPs only for small displacements and librations around the equilibrium configuration illustrated at the top of Fig.~\ref{Fig:Rod diagram}.} We consider two sizes of rods that are described in Table~\ref{Table:Rod Sizes}. The IRPs for the translational degrees of freedom in the smaller R1 rod \textcolor{black}{resemble the results for the point-like dipole in a standing wave trap \cite{maurer2022quantum}}. 
This is to be expected as the dimensions are sufficiently sub-wavelength. We also note a predicted information pattern for the $\theta_x$ \textcolor{black}{libration}. This is initially surprising as a cylindrical rod would display symmetry in this direction. We believe this to be due to a numerical artifact associated with the discrete geometry triangulation that does not display this exact symmetry. 

\begin{table}
    \centering
    \begin{tabular}{c c c}
        \hline \hline
        Rod Name & Diameter & Length\\
        \hline
        R1 & $0.108\mu\text{m}$ & $0.795\mu\text{m}$\\
        R2 & $1.55 \mu\text{m}$ & $9.024\mu\text{m}$\\
        \hline \hline
    \end{tabular}
    \caption{Dimensions of the rods considered in the text and in Fig. \ref{Fig:Rod diagram}. The R1 rods are based on the nanorods used in Ref.  \cite{Kuhn2015} while the R2 rods dimensions are arbitrarily chosen to have a diameter equal to the laser wavelength, and a length close to the beam waist to ensure we are no longer in the point-like regime.}
    \label{Table:Rod Sizes}
\end{table}



\begin{figure}
    \centering
    \includegraphics[width=0.48\textwidth]{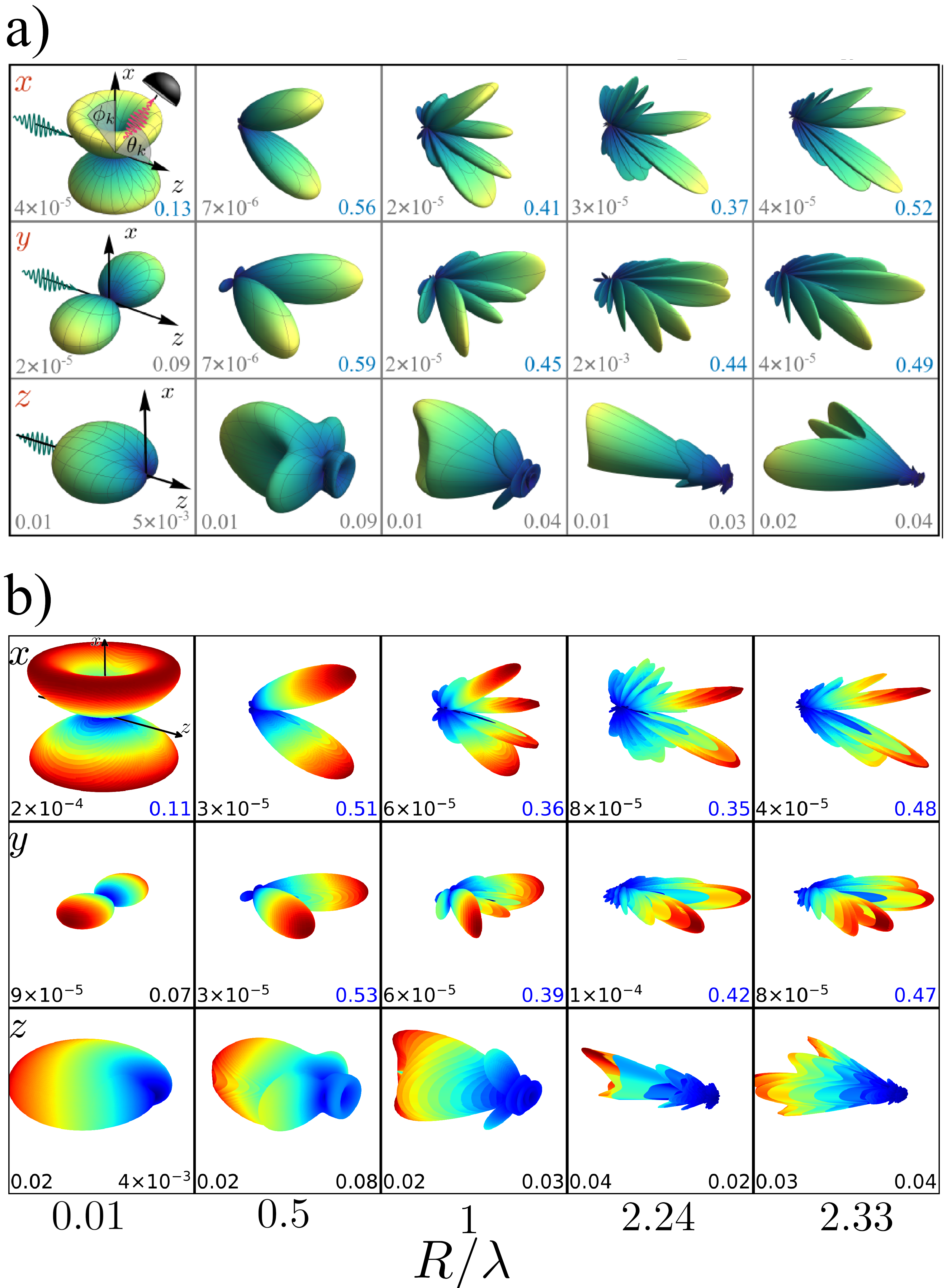}
     \caption{Information patterns for varying radii of spherical particles in a Gaussian beam traveling in the positive $z$ direction. The beam is focused by a lens to the left with a numerical aperture of $\text{NA}=0.1$. a) Shows the analytical results reproduced from~\cite{maurer2022quantum} and b) are the numerical results from SCUFF-EM. The numbers in the bottom corners of each panel refer to the detection efficiencies $\eta_\mu$ for a lens to the left and to the right.  The left collection lens is the same as the focusing lens, while the right collection lens has a numerical aperture of $\text{NA} = 0.75$.}
    \label{Fig:IRP Scuff Compare}
\end{figure}

\begin{figure}
    \centering
     \includegraphics[width=0.48\textwidth]{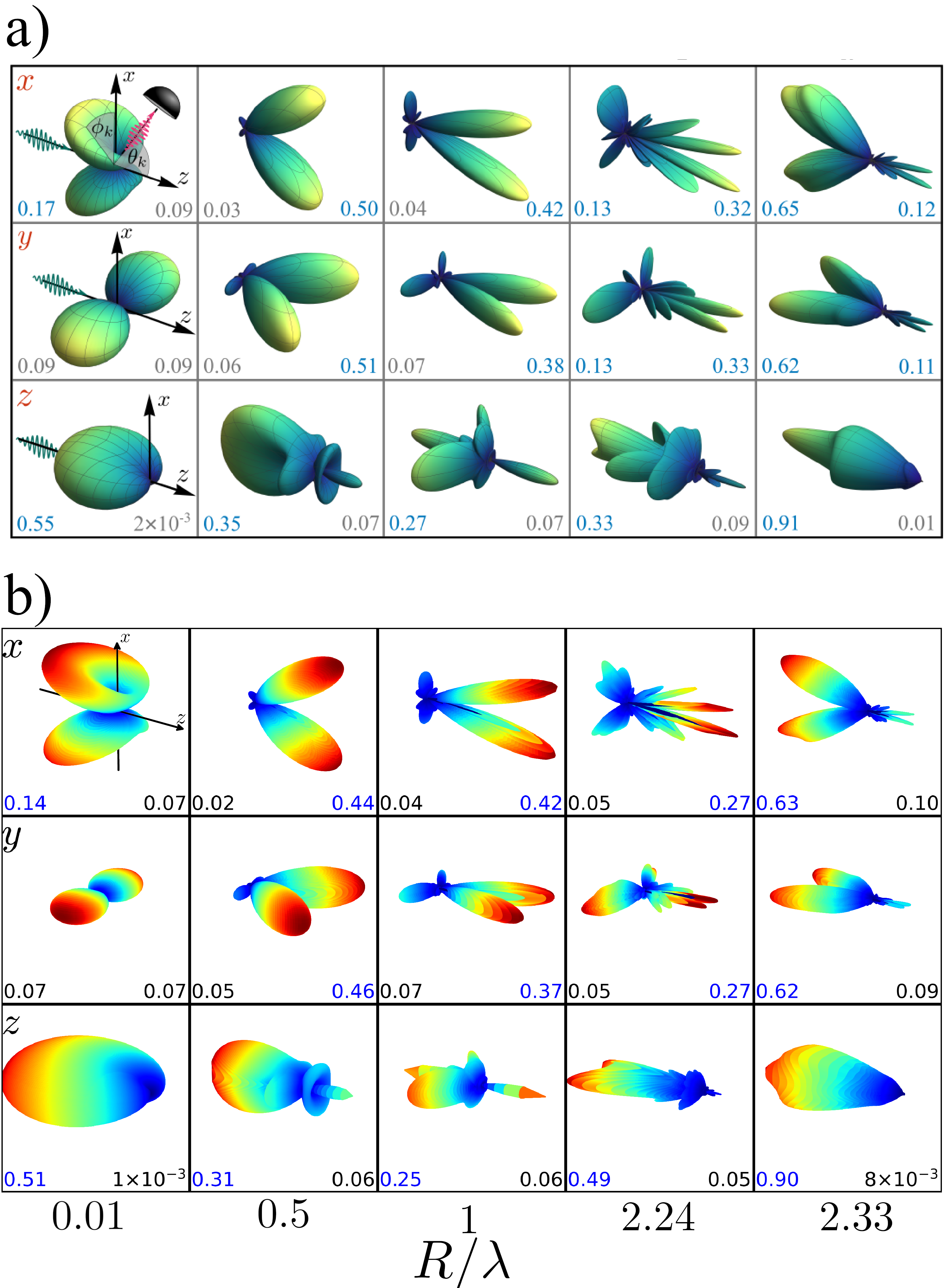}
    \caption{Information patterns for varying radii of spherical particles in a Gaussian beam traveling in the positive $z$ direction. The beam is focused by a lens to the left with a numerical aperture of $\text{NA}=0.75$ which is also used as the left collection lens for the detection efficiencies. The right collection lens also has a numerical aperture of $\text{NA}= 0.75$. a) shows the results reported from~\cite{maurer2022quantum} and b) are the numerical results from SCUFF-EM.}
    \label{Fig:IRP_Scuff_b}
\end{figure}

Finally, in order to validate our numerical method, we reproduce the results for spheres trapped in a single-beam trap as given in~\cite{maurer2022quantum,Maurer2024}. In Fig.~\ref{Fig:IRP Scuff Compare} and Fig. \ref{Fig:IRP_Scuff_b} we compare the theoretical results obtained with the method of Ref. \cite{maurer2022quantum} with numerical results as obtained from SCUFF-EM for IRPs for varying radii of spherical particles for different trapping and imaging numerical apertures. 
Reasonable agreement is obtained between the two methods, with small differences attributed to discretization of the spherical geometry in a finite mesh and numerical precision of the relevant derivatives needed to evaluate Eq. \ref{EQ:Fisher Flux}.

\section{Comparison with interferometry using a practical reference field
\label{sec:reference}}

In the previous section, we have determined the IRPs achievable for levitated objects of varying shape according to the theoretical limit from the Fisher information, subject to constraints from quantum radiation pressure back-action and shot noise.  To generate an interferometric detection of the particles motion in the laboratory, for example using homodyne detection, one must interfere the scattered light from the particle with another laser beam which is taken as a reference.  In this section, we compare the detection efficiency achievable for the example of a reference beam corresponding to one of the counter-propagating  trapping beams used in a dual beam optical trap in a recent experiment \cite{winstone2022optical}.

To ensure a strong interference term, the polarization of the reference beam should match that of the scattered field, and the reference field amplitude should be large enough such that the overall shot noise is dominated by the (bright) reference beam. Balanced photodetectors are often used in practice to cancel the non interferometric offset and suppress common mode laser-noise in \(\lvert \vec{E}_\mathrm{ref}(\vec{r})\rvert^2\), thereby reducing technical noise and improving displacement sensitivity.
In our simulations, we find the IRP with a reference field by taking the scattered field from a given back end, e.g. SCUFF-EM, and interfering it with the field from a single trapping beam in our analysis code. We then observe how the total \textcolor{black}{electromagnetic} field changes at the position $(\theta,\phi)$ as we change the degree of freedom we are interested in by an infinitesimal amount.

\begin{figure}
    \centering
    \includegraphics[width=0.25\textwidth]{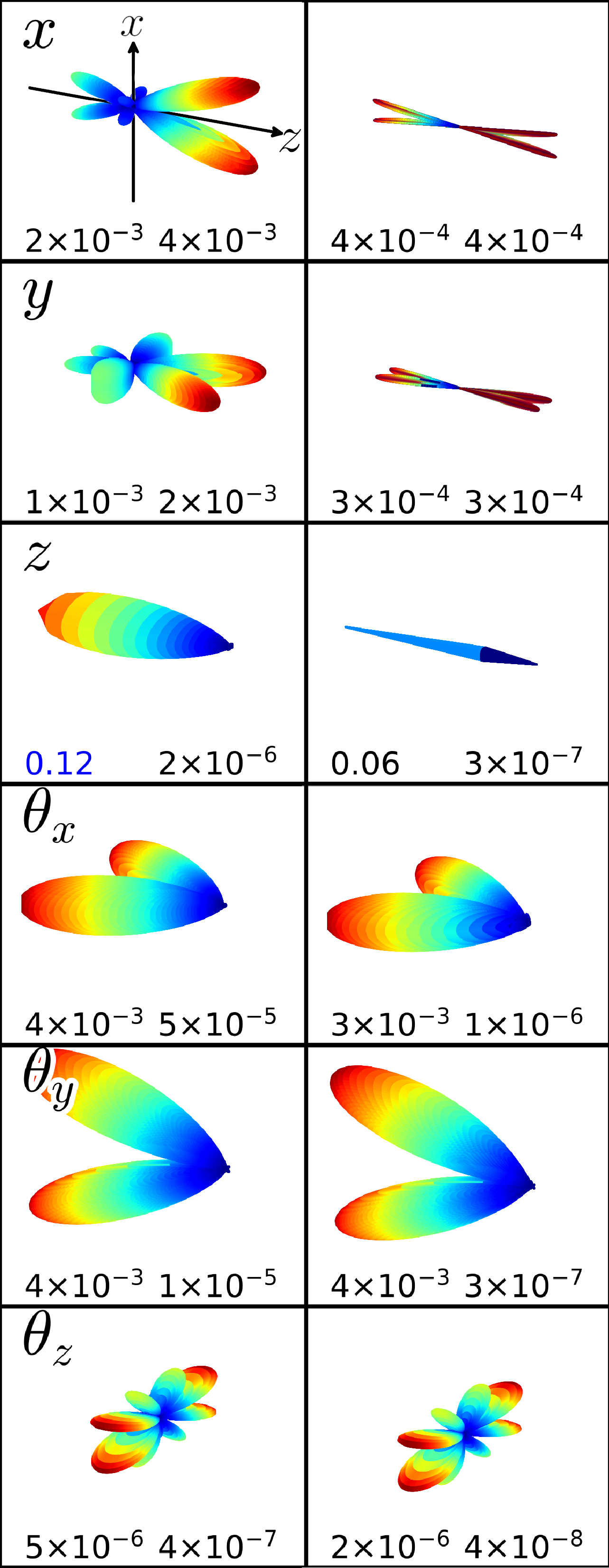}
    \caption{The information pattern for a H1 hexagonal plate illuminated by a single Gaussian beam with wavelength $\lambda = 1550   ~\text{nm}$ and a beam waist of $w_0 = 12 ~\mu\text{m}$. The left set of panels are the IRPs created using the Fisher information approach discussed in the main text. The panels on the right use a reference field to extract the information. This reference field is taken to be a Gaussian beam traveling in both the positive and negative $z$ directions. \textcolor{black}{For improved numerical precision, the reference beam has the same parameters as the trapping beam with a $10^5$ increase in magnitude and a $\pi/2$ phase shift used to maximize the amount of information that can be obtained from the phase of the scattered light. These IRPs are generated by interfering the scattered field with the reference and using Eq. \eqref{EQ:Scuff-Laser-Ref}}}
    \label{fig:Hex Laser Ref IRP}
\end{figure}

\textcolor{black}{The following expression for an experimental information flux estimator $\mathcal{I}_{\mu,i}^\text{exp}$ using a possibly non-ideal, practical reference field is derived in Appendix \ref{AppC} using the formalism in  \cite{Rotter2024}
\begin{equation}
    \mathcal{I}^\text{exp}_{\mu,i} = 
    \frac{ (\partial_\mu \text{Re}[\vec{E}^*(\theta,\phi,\mu)\times\vec{H}(\theta,\phi,\mu)] \cdot \vec{A}_i )^2 }
    {2\hbar \omega  \text{Re}[\vec{E}^*(\theta,\phi,\mu)\times\vec{H}(\theta,\phi,\mu)] \cdot \vec{A}_i},
    \label{EQ:Scuff-Laser-Ref}
\end{equation}
where the fields are 
\begin{equation}
    \begin{split}
        \vec{E}(\theta,\phi,\mu) &= \vec{E}_\text{scatt}(\theta,\phi,\mu) + \vec{E}_\text{ref}(\theta,\phi)\\
        \vec{H}(\theta,\phi,\mu) &= \vec{H}_\text{scatt}(\theta,\phi,\mu) + \vec{H}_\text{ref}(\theta,\phi).
        \label{EQ:scatt and ref field}
    \end{split}
\end{equation}}
\textcolor{black}{In order to find the detection efficiency analogous to  Eq~\eqref{EQ:Detection efficencies} except when using a practical reference field, we find the normalized information $\bar{\mathcal{I}}^\text{exp}_\mu(\theta,\phi)$ by dividing the experimental information by the total Fisher information, such that 
\begin{equation}
    \bar{\mathcal{I}}^\text{exp}_{\mu,i} = \frac{\mathcal{I}_{\mu,i}^\text{exp}}{\sum_{A_i} \mathcal{I}_{\mu,i}^\text{Fisher}}
    \label{EQ:Normed exp Info}
\end{equation}
}
\textcolor{black}{A comparison of the theoretical Fisher information radiation with the information scattered using a reference beam for the H1 hexagon is shown in Fig. \ref{fig:Hex Laser Ref IRP}, where we numerically evaluate this expression using finite differences. The reference beam is chosen to be a Gaussian beam traveling along both $z$ and $-z$, and the phase of the reference beam is chosen to maximize the information observable in the backwards direction, as illustrated in the third row of the right column of  Fig. \ref{fig:Hex Laser Ref IRP}. We note that for the case of the Fisher information, we here consider a single Gaussian beam and so the detection efficiencies reported for the left column differ from those in the first column of Fig. \ref{Fig:Hex diagram}, where the efficiencies in the forwards and backwards collection lens in Fig. \ref{Fig:Hex diagram} are approximately the average of the forwards and backwards efficiency reported in Fig. \ref{fig:Hex Laser Ref IRP} for illumination with a standing wave due to the symmetry of the pattern caused by the counter-propagating beams.
Here the collection lens corresponds to an effective numerical aperture of approximately $0.082$, and under these conditions, the detection of the translational degrees of freedom using the trap laser reference beam is at most 50 percent as efficient as the theoretical limit.  Nevertheless, this approach can be used to benchmark the sensitivity of practical interferometry setups relative to the best achievable. }







\section{Conclusion}


In this paper we have described a general method to numerically determine the information radiation patterns for dielectric particles of arbitrary shape based on a computation of the Fisher information. We have demonstrated the equivalence of the numerical results we obtain when using the Fisher information method and the analytic results for spherical Mie scattering described in Ref. \cite{maurer2022quantum,Maurer2024}. We have applied our method to commonly used non-spherical geometries in levitated optomechanics, including hexagonal high aspect ratio prisms and nanorods.  Due to its generalized approach, our method is broadly applicable to the optomechanics community for a wide range of trappable species commonly used in levitated experiments, as well as the wider quantum sensing, and possibly quantum computing community. Our approach also permits the calculation of the detection efficiency for interferometric detection with realistic reference laser beam configurations, which can be useful for experimental groups to determine the best imaging setups to achieve center of mass and \textcolor{black}{librational} cooling of trapped dielectric objects of non-spherical geometry. For example, such an approach can be beneficial for levitated optomechanical sensors which realize their ideal sensitivity by having a non-spherical shape, including rods or dumbbells for rotation sensing or high aspect ratio plate like particles for the application of high frequency gravitational wave detection above the frequency range of 10 kHz \cite{aggarwal2022searching,winstone2022optical}. 


\section*{Data Availability Statement}

The simulation code that supports the findings of this article is openly available~\cite{laing_2026_18379291}.

\section{Appendix} 

\subsection{Comparison of Numerical Methods\label{AppA}}


In this appendix, we provide a comparison between the results obtained using various computational back-ends described in Section \ref{sec:methods}, namely SCUFF-EM, PyGDM, and COMSOL for the H1 hexagon, to verify the general shape of the information radiation pattern when computed using different numerical solvers. The results displayed in Fig. \ref{fig:backendcompare} can be compared with the first column of plots displayed in Fig. \ref{Fig:Hex diagram}, but for a single beam rather than a dual beam trap. 
On the left, we show the 3D IRP plots generated using SCUFF-EM using the same methods as all previous figures. On the right, we show 2D cross-sections of the information pattern overlayed for each of the computational back-ends considered here. 
We observe reasonable qualitative agreement for all degrees of freedom shown. 


\begin{figure}
    \centering
    \includegraphics[width=0.5\textwidth]{BackendCompare_FullFigurev9_highres.png}
    \caption{Information radiation pattern comparison for a hexagon in a single beam trap for various computational backends. (left) 3D IRP plots from the SCUFF-EM implementation for each degree of freedom $\mu$ of an H1 hexagonal disk with dimensions given in Table~\ref{Table:Hex Sizes} trapped in a single Gaussian beam linearly polarized along the $x-$ direction and propagating along the positive $z-$ direction with a wavelength of $\lambda = 1550~\text{nm}$ and a beam waist of $w_0 = 12~\mu\text{m}$. (right) 2D cross-sections of the IRPs in the $x=0$, $y=0$, and $z=0$ planes for each of the backends discussed in Section \ref{sec:methods}, SCUFF-EM, PyGDM, and COMSOL.}
    \label{fig:backendcompare}
\end{figure}

\begin{figure}
    \centering
    \includegraphics[width=0.5\textwidth]{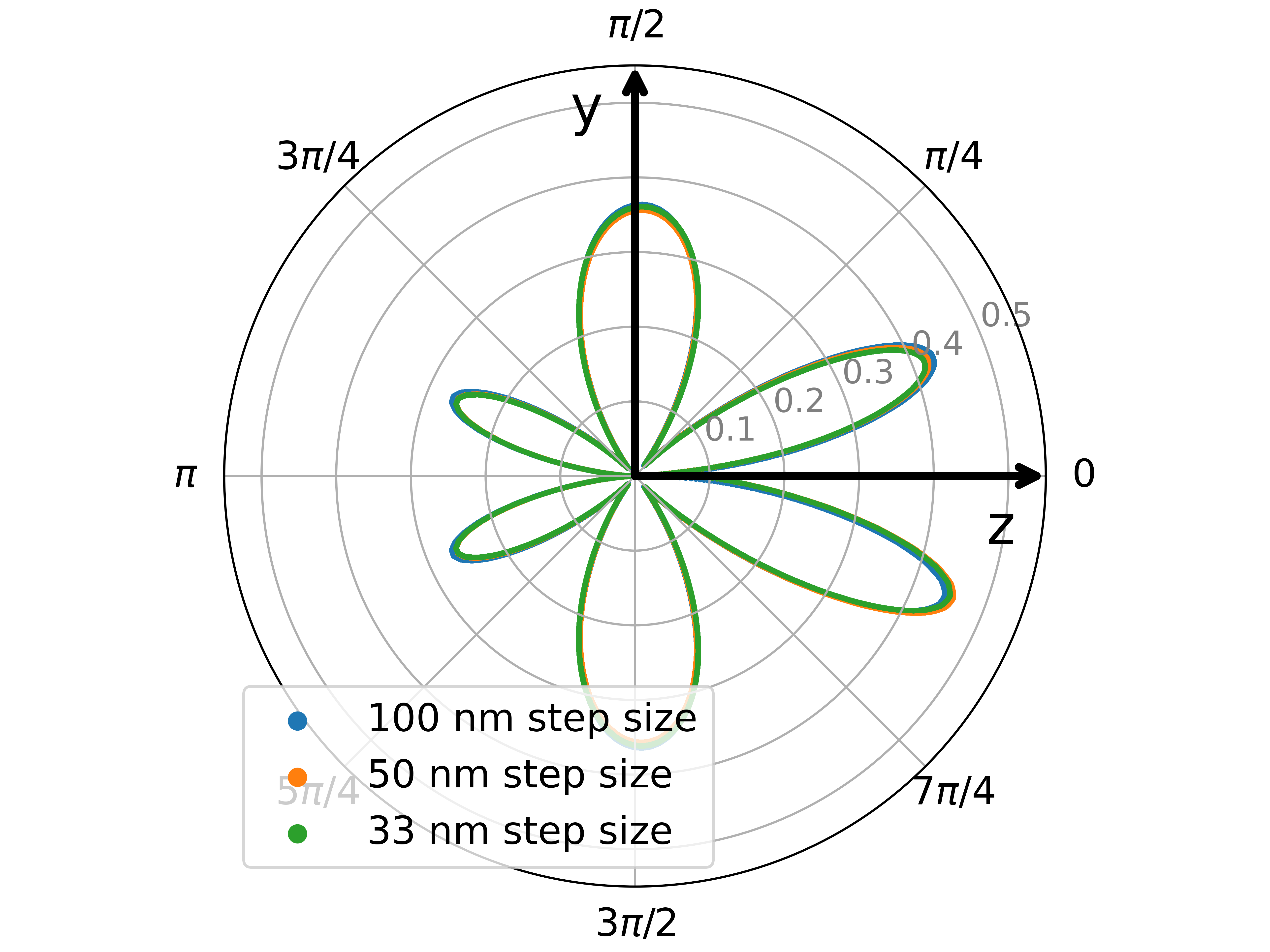}
    \caption{Information radiation pattern comparison for a hexagon in a single beam trap for varied mesh discretizations computed in PyGDM2. A 2D cross-section of the IRP for the y motion in the x=0 plane was taken for a similar case as in Fig. \ref{fig:backendcompare} where the step size of the cubic mesh was varied from 100 nm to 33 nm.}
    \label{fig:pygdmstepcompare}
\end{figure}

\textcolor{black}{We also explored convergence of the individual computational back-ends. Fig. \ref{fig:pygdmstepcompare} shows an example of convergence of the PyGDM2 code for the case of a hexagon in a single beam trap, similar to the case shown in Fig. \ref{fig:backendcompare}. The step size of the cubic mesh discretization was varied to 100 nm, 50 nm, and 33 nm, which correspond to 1167, 9546, and 33034 total mesh elements. While there were slight variations in the result due to numerical error, the result is converged to a solution. Similar verifications of convergence were done for each computational back-end.}

\subsection{Pattern Symmetry Analysis\label{AppB}}

\textcolor{black}{In Section \ref{sec:results}, we discussed the symmetries of the information patterns that arise from the symmetries of the geometry of the particles and trap beams. In addition to the observations made in the text, there are several studies that could be done to tease out the source of certain features in the patterns. One such feature that we choose to highlight here is the number of radial lobes for the $\theta_z$ motion. }

\begin{figure}
    \centering
    \includegraphics[width=0.5\textwidth]{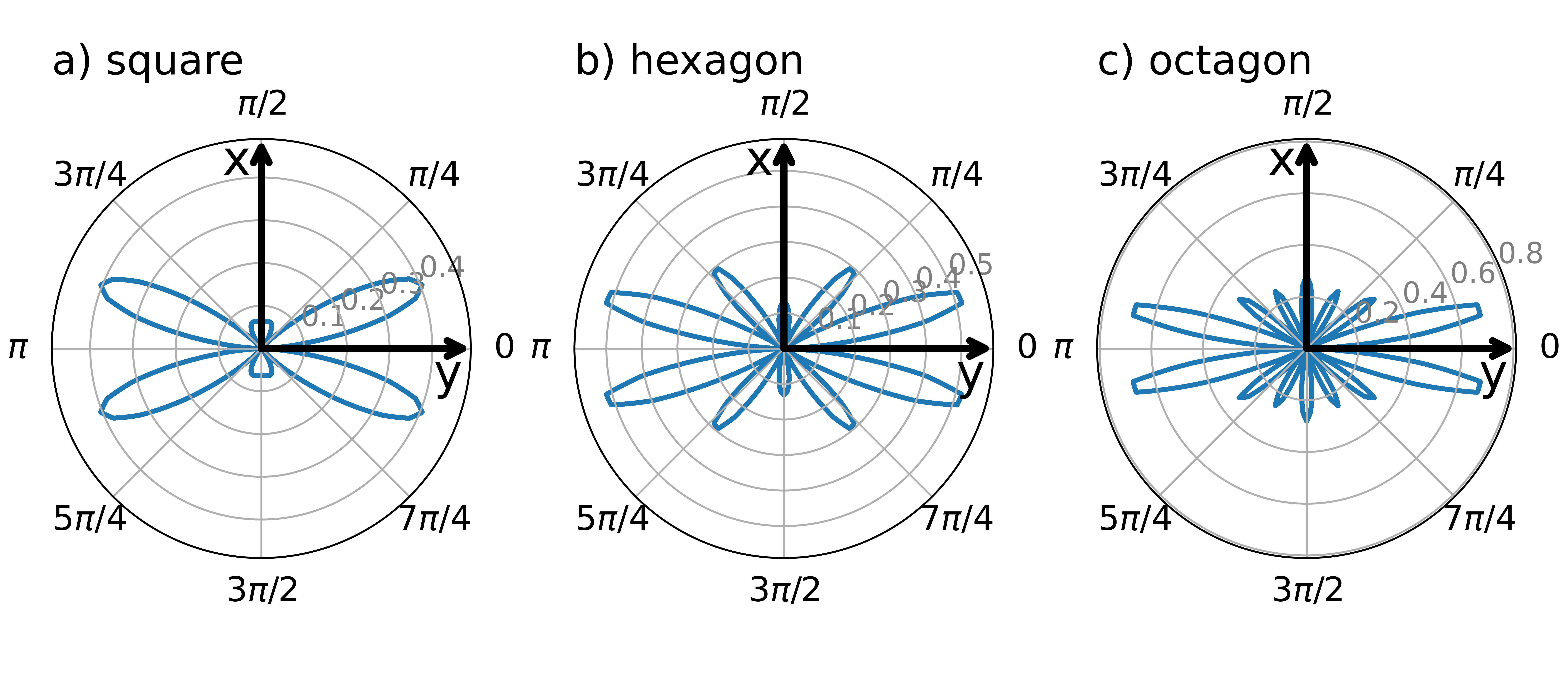}
    \caption{\textcolor{black}{Comparison of information pattern lobes for the $\theta_z$ motion for polygons of varied number of sides, namely (a) a square, (b) a hexagon, and (c) an octagon. These patterns are computed for plate-like particles in a single beam trap linearly polarized along the $x-$ direction and propagating along the positive $z-$ direction with a wavelength of $\lambda = 1550~\text{nm}$ and a beam waist of $w_0 = 12~\mu\text{m}$. The particles are chosen to have the same thickness and point-to-point diameter as the H1 hexagon. The simulations are performed in PyGDM, and the patterns are plotted as 2D cross-sections in the $z=0$ plane.}}
    \label{fig:symmetry}
\end{figure}

\textcolor{black}{
We performed simulations of the information patterns for plate-like particles where the face is a polygon with a varied number of sides, namely a square, hexagon, and octagon. The particles have the same thickness and point-to-point diameter as the H1 hexagon, and are trapped in a single Gaussian beam. The results are shown in Fig. \ref{fig:symmetry}. We observe that the number of lobes for these particles are $6$, $10$, and $14$, respectively, following a $N_{\mathrm{lobe}}=2(n_{\mathrm{side}}-1)$ pattern. This calculation shows that in certain cases the observed number of lobes depends on the symmetry of the trapped object.}
 
 \textcolor{black}{We note that in general the number of lobes appearing in the IRPs also depends on the size of the object compared to the wavelength, independently of the symmetry of the object. For instance, for spherical objects of radius $R$, as $R/\lambda$ increases, we observe a pattern of lobes evolving due to Mie resonances (see e.g. Ref. \cite{maurer2022quantum} and Figs. \ref{Fig:IRP Scuff Compare} and \ref{Fig:IRP_Scuff_b}).
Further studies could be done to gather a better understanding of the source of all the lobes and symmetries that appear in the many patterns shown in this work, but would require further analysis beyond the scope of the present paper and is reserved for future investigations.
}

\subsection{Information Radiation Patterns for Non-Ideal Reference Fields}
\label{AppC}

\textcolor{black}{In this section, we follow the formalism provided in \cite{Rotter2024} to find the information extracted via homodyne measurement.  In our description, we assume the light is described by coherent states and therefore exhibits Poissonian photon statistics. A description of, for example, squeezed light, would require a more rigorous quantum mechanical treatment.
For a photon flux $\Phi_i$ through an area $A_i$, the information contained in the photon flux about the degree of freedom $\mu$ is
\begin{equation}
\begin{split}
    \mathcal{I}_{\mu,i} &= \frac{(\partial_\mu \Phi_i(\mu))^2}{\Phi_i(\mu)}\\
    &= \frac{ [\partial_\mu \vec{S}(\theta,\phi,\mu) \cdot \vec{A}_i ]^2 }{\hbar \omega \vec{S}(\theta,\phi,\mu) \cdot \vec{A}_i )},
\end{split}
\label{EQ:Poisson Info}
\end{equation}
where we have used the time-averaged Poynting vector
\begin{equation}
    \vec{S} = \frac{1}{2}\text{Re}[\vec{E}^*(\theta,\phi,\mu)\times\vec{H}(\theta,\phi,\mu)]
    \label{EQ:TimeAve Poynting}
\end{equation}
and can thus rewrite Eq~\eqref{EQ:Poisson Info} as
\begin{equation}
    \mathcal{I}_{\mu,i} = 
    \frac{ (\partial_\mu \text{Re}[\vec{E}^*(\theta,\phi,\mu)\times\vec{H}(\theta,\phi,\mu)] \cdot \vec{A}_i )^2 }
    {2\hbar \omega \text{Re}[\vec{E}^*(\theta,\phi,\mu)\times\vec{H}(\theta,\phi,\mu)] \cdot \vec{A}_i}
    \label{EQ:RefField Info},
\end{equation}
    which is Eq~\eqref{EQ:Scuff-Laser-Ref} in the main text.  By using the Cauchy Schwartz inequality, this is less than or equal to the Fisher information, as shown in Ref. \cite{Rotter2024}.
We can continue to break Eq~\eqref{EQ:RefField Info} into its scattered and reference components by inserting Eq~\eqref{EQ:scatt and ref field} into Eq~\eqref{EQ:TimeAve Poynting}, we get
\begin{equation}
    \begin{split}
        \vec{S} =
        \frac{1}{2}
        \text{Re}[&(\vec{E}^*_\text{scat}+\vec{E}^*_\text{ref})\times (\vec{H}_\text{scat}+\vec{H}_\text{ref})]\\
       = \frac{1}{2}
        \text{Re}
        [&\vec{E}^*_\text{scat} \times \vec{H}_\text{scat} +
        \vec{E}^*_\text{scat}  \times \vec{H}_\text{ref}  +\\
        &\vec{E}^*_\text{ref}   \times \vec{H}_\text{scat} +
        \vec{E}^*_\text{ref}   \times \vec{H}_\text{ref}  ].
        \end{split}
        \label{EQ:Split out Poynting}
\end{equation}
Using the product rule, and the fact that the reference field does not depend on the degree of freedom $\mu$, we find the differential of the Poynting vector to be
\begin{equation}
    \begin{split}
        \partial_\mu \vec{S}=\frac{1}{2} \text{Re}[
        \partial_\mu\vec{E}^*_\text{scat} &\times \vec{H}_\text{scat} +
        \vec{E}^*_\text{scat} \times \partial_\mu \vec{H}_\text{scat} +\\
        \partial_\mu\vec{E}^*_\text{scat}  &\times \vec{H}_\text{ref} +\vec{E}^*_\text{ref}   \times \partial \vec{H}_\text{scat} ].
    \end{split}
    \label{EQ:Poynting derivative}
\end{equation}
Inserting Eq~\eqref{EQ:Poynting derivative} into Eq~\eqref{EQ:Poisson Info}, we find the total information contained in the total field to be
\begin{widetext}
\begin{equation}
    \mathcal{I}_{\mu,i} = \frac
    {(\text{Re}[
        \partial_\mu\vec{E}^*_\text{scat} \times \vec{H}_\text{scat} +
        \vec{E}^*_\text{scat} \times \partial_\mu \vec{H}_\text{scat} +
        \partial_\mu\vec{E}^*_\text{scat}  \times \vec{H}_\text{ref} +
        \vec{E}^*_\text{ref}   \times \partial_\mu \vec{H}_\text{scat} ] \cdot \vec{A}_i)^2}
    {2\hbar\omega\text{Re}
        [\vec{E}^*_\text{scat} \times \vec{H}_\text{scat} +
        \vec{E}^*_\text{scat}  \times \vec{H}_\text{ref}  +
        \vec{E}^*_\text{ref}   \times \vec{H}_\text{scat} +
        \vec{E}^*_\text{ref}   \times \vec{H}_\text{ref}  ]\cdot \vec{A}_i}
\end{equation}
\end{widetext}
}


\section*{Acknowledgements}
We are grateful to Oriol Romero-Isart for helpful discussions which contributed significantly to the development of this work. We thank Patrick Maurer for assistance with comparison of our results with Lorentz-Mie scattering by dielectric spheres. AAG is supported in part by NSF grants PHY-2110524 and PHY-2111544, the Heising-Simons Foundation, the John Templeton Foundation, and ONR Grant N00014-18-1-2370. AAG and GW are supported by the W.M. Keck Foundation. A.G.~also acknowledges support from NSF grants PHY-2409472, DARPA, the Gordon and Betty Moore Foundation Grant GBMF12328, DOI 10.37807/GBMF12328, the Alfred P.~Sloan Foundation under Grant No.~G-2023-21130, and the Simons Foundation. AG is supported by the Alexander von Humboldt foundation.
SL is also supported by EPSRC International Quantum Technologies Network Grant EP/W02683X/1 and is grateful for EPSRC support through Standard Research Studentship (DTP) EP/R51312X/1. 
This research was supported in part through the computational resources and staff contributions provided for the Quest high performance computing facility at Northwestern University which is jointly supported by the Office of the Provost, the Office for Research, and Northwestern University Information Technology.


\bibliography{apssamp}

\end{document}